\documentclass[]{aa}
\usepackage[T1]{fontenc}
\usepackage[utf8]{inputenc}

\usepackage{graphicx}
\usepackage{natbib}
\usepackage{ulem}
\usepackage{textcomp}
\usepackage{gensymb}
\usepackage{longtable}
\usepackage{threeparttable}
\usepackage{multicol}
\usepackage{multirow}
\usepackage{float}
\usepackage{todonotes}
\usepackage{footnote}
\usepackage[Symbol]{upgreek}
\usepackage{amsmath}
\usepackage{amssymb}
\usepackage{comment}
\usepackage{etoolbox}
\usepackage{txfonts}
\usepackage{url}
\usepackage{xcolor}

\usepackage[most]{tcolorbox}
\usepackage{hyperref}
\hypersetup{
     colorlinks=true,
     linkcolor=blue,
     filecolor=blue,
     citecolor = blue,      
     urlcolor=blue,
     }
\usepackage{xcolor}

\newcommand{\erosita}{{\small eROSITA}}
\newcommand{\apec}{{\small APEC} }
\newcommand{\disnht}{{\tt disnht}}
\newcommand{\xspec}{{\small XSPEC}}
\newcommand{\tbabs}{{\tt tbabs}}

\begin{document}

\title{disnht: modeling X-ray absorption from distributed column densities}

\author{Nicola Locatelli\inst{1}, Gabriele Ponti\inst{2,1}, Stefano Bianchi\inst{3} }

\offprints{%
 E-mail: nlocat@mpe.mpg.de}
\institute{Max-Planck-Institut f\"ur Extraterrestrische Physik (MPE), Giessenbachstrasse 1, 85748 Garching bei M\"unchen, Germany
\and INAF - Osservatorio Astronomico di Brera, via E. Bianchi 46, 23807 Merate (LC), Italy
\and Dipartimento di Matematica e Fisica, Universit\'a degli Studi Roma Tre, via della Vasca Navale 84, I-00146 Roma Italy}

\authorrunning{N. Locatelli et al.}
\titlerunning{absorption from distribution of column densities}

\date{Accepted ???. Received ???; in original form ???}

\abstract{
Collecting and analysing X-ray photons over either spatial or temporal scales encompassing varying optical depth values requires knowledge about the optical depth distribution. In the case of sufficiently broad optical depth distribution, assuming a single column density value leads to a misleading interpretation of the source emission properties, nominally its spectral model.
We present a model description for the interstellar medium absorption in X-ray spectra at moderate energy resolution, extracted over spatial or temporal regions encompassing a set of independent column densities. The  absorption model (named \disnht) approximates the distribution with a lognormal one and is presented in table format. The solution table and source code are made available and can be further generalized or tailored for arbitrary optical depth distributions encompassed by the extraction region. The \disnht\, absorption model presented and its generalized solution are expected to be relevant for present and upcoming large angular scale analyses of diffuse X-ray emission, such as the ones from the extended ROentgen Survey with an Imaging Telescope Array (\erosita) and the future {\it Athena} missions. } 
\maketitle

\label{firstpage}

\section{Introduction}

Neutral and partially ionized material can efficiently absorb ultra-violet light and X-rays \citep{2000ApJ...542..914W}. Absorption is observed as an exponential decrease in the received intensity of the light $I\propto e^{-\tau}$, quantified through the optical depth $\tau= \sigma N_H$, where $N_H$ is the integral of its neutral hydrogen along the line of sight in cm$^{-2}$. The abundances of all the other metals contributing to the absorption in the intervening medium are also taken into account through the absorption cross section $\sigma( X(Z), E )$, in cm$^2$, as a function of the photon energy $E = h\nu$, where $X(Z)$ maps the relative abundance of the element Z with respect to the hydrogen column density $N_H$ \citep{1982AN....303..142R}.
The interstellar medium (ISM) hosts large amounts of such an absorbing material in the form of a complex multiphase collection of ions, atoms, molecules and dust grains diffused all across the thin stellar disk of our Galaxy, encompassing a wide range of densities ($10^{-4} < n_{H,{\rm ISM}} < 10^6\, \rm cm^{-3}$) and temperatures ($T_{\rm ISM} < 10^6 K$) and contributing differently to the X-ray absorption \citep[e.g.][for a review]{2003ApJ...587..278W, 2001RvMP...73.1031F}.
When analysing the light spectrum along one line of sight the absorption can be easily detected and accounted for, thanks to its known characteristic energy dependence (namely well known cross sections, \citealt{1992ApJ...400..699B}). However, if the light collected into an instrument and summed into a spectrum encompasses several independent $N_H$ elements (in space or time), an appropriate description of the absorption necessarily requires a description of the $N_H$ distribution of values, as a single column density value describing the absorption within a given (and satisfactory) accuracy may simply not exist.
This fact follows from the mathematical evidence that a sum over exponential functions can be approximated by a single exponential function only in the case of a very peaked distribution of exponents around one value.
An incomplete but instructive list of cases in which this assumption is broken include the spectral analysis of: 
faint diffuse emission for which collection of photons over a large angular scale is performed to increased the significance of the spectral features \citep[e.g.][]{2009ApJS..181..110M, 2009PASJ...61S.115M, 2009PASJ...61..805Y, 2012ApJ...756L...8G, 2015MNRAS.453..172P, 2021MNRAS.504.1627C, 2021arXiv211006995W}; 
extraction regions holding high ($\sim 10^{24}$ cm$^{-2}$) and very high ($\sim 10^{25}$ cm$^{-2}$) absorption such as toward dense clouds in the Galactic disc, for which collecting the photons over even a few resolution elements can significantly affect the X-ray spectrum \citep[e.g.][]{2011ApJ...735L..33M};
pointlike sources holding column densities variable over the observation timescale, such as for dipping low mass X-ray binaries binaries \citep{2006A&A...445..179D, 2016AN....337..512P, 2017MNRAS.472.2454B}, changing-look active galactic nuclei \citep{2003MNRAS.342..422M, 2007MNRAS.377..607P, 2009ApJ...695..781B, 2009ApJ...696..160R, 2012MNRAS.421.1803M, 2015ApJ...800..144L} and magnetic accreting white dwarfs \citep{1998MNRAS.298..737D}.
Modeling of the spectral absorption through distribution of column densities has already successfully performed in the past using a power law column density distribution \citep[{\tt pwab} model, presented in ][]{1998MNRAS.298..737D} and more recently using a lognormal distribution \citep{2021MNRAS.504.1627C, 2021arXiv211006995W}. These examples show the increasing need in the X-ray community of careful and systematic treatment of the absorption when its properties change in a non-negligible way across space or time.
For these and all the other cases mentioned above, in this work we aim to provide the community with a method and precomputed tables of absorption factors $-\ln \,A_\Theta (E)$ able to properly reproduce the absorption curve. The model produced will be useful in general to any X-ray spectral analysis that involves a set of independent column densities broadly distributed around an expected value. The DIStributed NH Table model (\disnht) is written into an {\tt ascii} readable table and is also already implemented in .fits format readable from the X-ray SPECtral fitting software \xspec\, as an exponential table model (etable).
All the tables and materials used to create them, including the source code are publicly available \footnote{\url{http://www.brera.inaf.it/hotmilk/models.html}}. The code can be used to compute the absorption spectrum for any arbitrary column density distribution.
In Section~\ref{sec:model} we describe the model creation and output table format and usage. We present different realistic scenarios in Sec.~\ref{sec:discussion} and we draw our conclusions in Sec.~\ref{sec:conclusion}.

\section{Model} \label{sec:model}

For any given set of M column densities $\Theta_M(N_H)$, the exact overall absorption spectrum can be computed as
\begin{equation}
    A_\Theta = \sum_i^M \exp\left\{ -\sigma(X(Z), E) \, N_{H,i} \right\} \label{eq:abs_exact}
\end{equation}
where the cross section $\sigma$ is derived from \cite{1992ApJ...400..699B} and assuming \citealt{2003ApJ...591.1220L} as our benchmark model for the metals in the ISM. Cross sections from different abundance sets commonly used in the \xspec\, software are also provided for completeness \citep{1989GeCoA..53..197A, 2009ARA&A..47..481A}.

\begin{figure*}
    \centering
    \includegraphics[width = 0.8\linewidth]{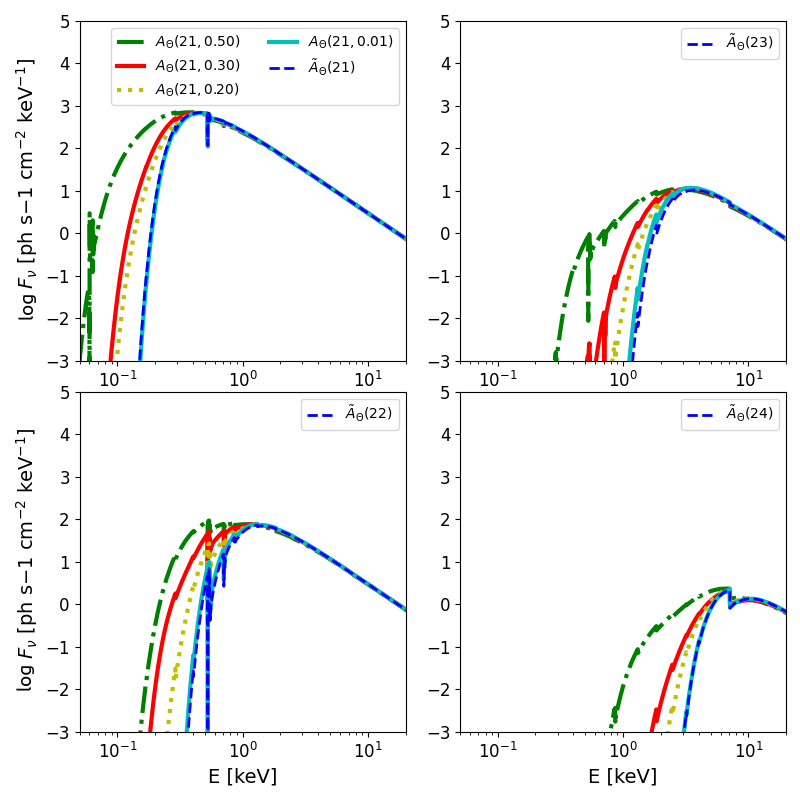}
    \caption{ Power law spectrum $F_\nu \propto \nu^{-2}$ absorbed by a distribution of column densities. The distribution $\Theta$ is drawn from a lognormal with avg$_\Theta = 21.0$ (upper left panel), 22.0 (upper right), 23.0 (lower left) and 24.0 (lower right) dex [cm$^{-2}$]. Distributions for different std$_\Theta$ are also shown in each panel: std$_\Theta= 0.5$ (green dotted dashed line), 0.3 (red solid), 0.2 (yellow dotted), 0.01 (cyan solid). 
    The blue dashed line show an absorbed flat spectrum $\tilde{A}_\Theta$ with same normalization, absorbed by $N_H = \rm avg_\Theta$ in each plot. 
    }
    \label{fig:NH_distr_25_21.0_0.2}
\end{figure*}

In the lack of an exact model for the absorption by a distribution of column densities, it is common to simply assume an average value $\langle N_H\rangle$ or to fit for a single value, so that the absorption spectrum is
\begin{equation}
    \tilde{A}_\Theta = \exp\left\{ -\sigma(X(Z), E) \, \langle N_H \rangle \right\} \label{eq:abs_avg}
\end{equation}
The discrepancy introduced in $\tilde{A}_\Theta$ with respect to the exact absorption spectrum $A_\Theta$ can be significant. We show an example in Fig.~\ref{fig:NH_distr_25_21.0_0.2}, where we accumulate the emission from a sample of absorbed spectra $p_{m=1 ... M}$ with $F_{\nu,i} \propto \nu^{-2}$ and fixed (arbitrary) normalization.
We assume the distribution of column density $\Theta_M(N_H)$ to be lognormal with expected value avg$_\Theta = 21, \, 22,\, 23,\, 24$ dex and standard deviation of std$_\Theta = 0.01,\, 0.2,\, 0.3,\,0.5$ dex. Every spectrum $p_m$ is absorbed by a column density in the set $\Theta_M$ opacity equal to the one used by the \tbabs\, model \citep{2000ApJ...542..914W}. The absorbed spectra are then summed together. The sum spectrum $A_\Theta$ (Eq.~\ref{eq:abs_exact}) is then compared with a single spectrum absorbed by the average value of $\Theta_M$, $\tilde{A}_\Theta$ (Eq.~\ref{eq:abs_avg}). 
We note that assuming a shape for the column density distribution $\Theta_M$ is necessary to make the data fit to converge. An arbitrary distribution can instead be modeled whenever precisely know, regardless of its shape.
We find that a lognormal distribution for the column density is a reasonable assumption in real cases where the discrepancy between the exact and average-value solutions for the absorbed spectrum becomes relevant (see Sec.~\ref{sec:discussion} for further details).
The lognormal distribution is parametrized by its expected value avg$_\Theta$ and standard deviation std$_\Theta$.

The discrepancy between the exact and average column density (blue dashed line in Fig.~\ref{fig:NH_distr_25_21.0_0.2}) solutions can be computed in terms of the ratio $A_\Theta / \tilde{A}_\Theta$. In the example shown in the upper left panel $A_\Theta / \tilde{A}_\Theta$ is in excess of $1\%$ at $E \leq 1$ keV, amounting to $10\%$ at 0.3 keV. The discrepancy increases exponentially at lower energies, so that it is already a few orders of magnitudes at 0.1 keV. This situation worsens for all values of avg$_\Theta > 21.0$ and std$_\Theta >> 0.01$ dex. 

We compute and tabulate the correct absorption $A_\Theta$ for all values of avg$_\Theta \in [19.5, 25.5]$ and std$_\Theta \in [0, 1]$. The table is multiplicative so that its entries, as a function of energy, are to be multiplied to any unabsorbed spectrum. The \xspec\, table provides instead the quantity $ - \ln A_\Theta$ so that, for instance, for $A_\Theta = e^{-2}$ the \xspec\, entry provides the value 2. The \xspec\, class of this table is an {\tt etable} and it also can be used as a multiplicative model when loaded through the command {\tt etable\{<TABLE\_NAME>\}}.

\section{Discussion} \label{sec:discussion}



%
\begin{figure*}
    \centering
    \includegraphics[width=0.7\linewidth]{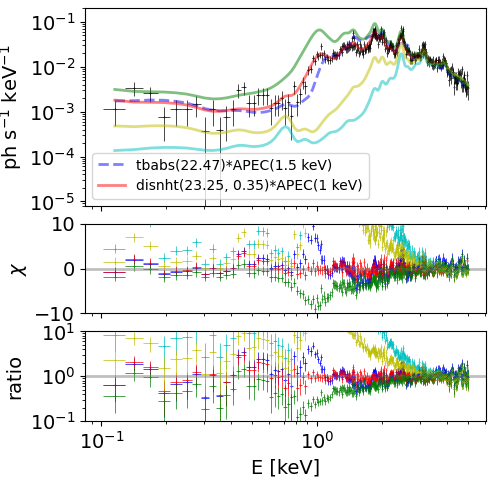}
    \caption{Simulated spectrum absorbed by a lognormal distribution of column densities. The emission model is an \apec\, at $T=1$~keV and solar abundances \citep{2003ApJ...591.1220L}. The mock data are the sum of 100 simulated \apec\, absorbed by different column density values of the neutral material extracted from a lognormal distribution. The solid lines show \disnht\, models with avg$_\Theta$ values: 23.25 (best-fit) and  different std$_\Theta$ 0.50 (green), 0.33 (red, best-fit), 0.20 (yellow) and 0.01 (cyan). The blue dashed line show the same emission model absorbed by a single column density equal to avg$_\Theta=23.25$ (with best-fit kT=1.5 keV). Residuals $\chi \equiv \rm (data - model)/error$ and ratios $\rm data / model$ are shown in the central and lower panel respectively.}
    \label{fig:disnhtApec_23.3_0.01-0.20-0.33-0.50_1_1_0.5_plot}
\end{figure*}
We demonstrate the efficacy of the \disnht\, model in retrieving the correct values by simulating a set of 100 lines of sight encompassing a distribution of column densities $N_H$. The $N_H$ distribution $\Theta$ is drawn from a lognormal with mean value avg$_\Theta = 23.30$ and standard deviation std$_\Theta = 0.35$. The avg$_\Theta$ and std$_\Theta$ recomputed after the extraction are 23.25 and 0.34 dex respectively.
For each $N_H$, we generate a mock spectrum using the {\tt fakeit} tool of \xspec. We build the spectrum from a collisionally ionized plasma model (\apec hereafter, \citealt{2001ApJ...556L..91S}) at a temperature $T_{\apec}=1$ keV and ISM metal abundances \citep[taken from][]{2003ApJ...591.1220L}, absorbed by neutral material, aiming to emulate the observation of a large region of diffuse emission from the hot phase of the ISM in the midplane of the Milky Way. We set the exposure time of each observation to 1 ks. The instrumental response matrix and ancillary files have been taken from XMM-{\it Newton} observations of the Galactic center region with the EPIC-PN camera \citep{2015MNRAS.453..172P}. After simulating each spectrum individually, we sum them. The resulting spectrum is shown in Fig.~\ref{fig:disnhtApec_23.3_0.01-0.20-0.33-0.50_1_1_0.5_plot}.
\begin{table}[]
    \centering
    \begin{tabular}{c | c | c c}
        comp & param & \tbabs & \disnht \\
        \hline
        absorption & $N_H$, avg$_\Theta$ & $3.29 \pm 0.06 \times 10^{22}$  & $23.27 \pm 0.03$ \\
        absorption & std$_\Theta$        &           -                   & $0.340 \pm 0.002$ \\
        \apec   & $kT$              & $1.77 \pm 0.04$        & $1.04 \pm 0.02$ \\
        \apec   & $Z / Z_\odot$     & $1.7 \pm 0.2$          & $1.13 \pm 0.15$ \\
        \apec   & norm              & $8.7 \pm 0.7 \cdot 10^{-4}$& $8.1 \pm 1.1 \cdot 10^{-3}$ \\
        \hline 
                & C stat            & 1594.4                 & 964.33 \\
                & d.o.f. $^\ddagger$& 976                    & 975 \\
    \end{tabular}
    \caption{ Best fit parameters from the spectral fit of the mock data. $\ddagger$ degrees of freedom. }
    \label{tab:fit_mockSum}
\end{table}
The best-fit values of the \disnht $\times $\apec\, model are reported in Tab.~\ref{tab:fit_mockSum}. Both the simulated distribution of column densities and the emission model are reproduced by the \disnht\, model best-fit parameters within the statistical uncertainties, holding a C statistics of $\sim 964$ over 975 degrees of freedom. The \tbabs $\times$ \apec\, best-fit model instead provides a column density of the absorbing material \citep{2000ApJ...542..914W} of $3.29 \times 10^{22} \,\rm cm^{-2}$, about an order of magnitude lower than the mean of the simulated (log-)distribution. Furthermore, when the \disnht\, model is used, the retrieved temperature and metallicity of the emission component are consistent with a $kT=1$ keV collisionally ionized plasma with solar abundances, whereas by using the \tbabs\, model, the plasma characterization ($kT=1.56 \pm 0.03$ keV, $Z = 1.7 \pm 0.2 \, Z_\odot$) significantly differs from the one originally set. The C statistics held by the best-fit \tbabs*\apec\, model is 1594.4 over 976 degrees of freedom. 
There is thus an overall improvement of the \disnht\, model with respect to \tbabs, both in terms of describing the absorption and (consequently) the emission properties of the spectrum, and of the statistical goodness of the fit, at the expense of an additional degree of freedom.

\subsection{Lognormal distribution approximation}

The \disnht\, model is also already provided and precomputed as an interpolation between the discrete entries of a two dimensional table. Each dimension of this table corresponds to one degree of freedom of the model. The physical distribution of column density in space or time can be arbitrarily complex to model, in terms of number of parameters required to describe it analytically. The (log-)normal distribution, if applicable, offers the advantage of providing physical insight rather than just a mathematical description and also to require a minimal amount of free parameters: a mean and a dispersion value. 
Is a (log-)normal distribution a sufficiently good representation of the column density distribution? 
We consider the assumption as sufficiently good by testing different observed column density distributions against a lognormal with the same mean, standard deviation and number of elements as the real distribution, through a Kolmogorov-Smirnov test. The approximation is valid for rejection probabilities of the null hypothesis (of the two distributions to have been drawn by the same one) larger than 0.05.
\begin{figure}
    \centering
    \includegraphics[width=\linewidth]{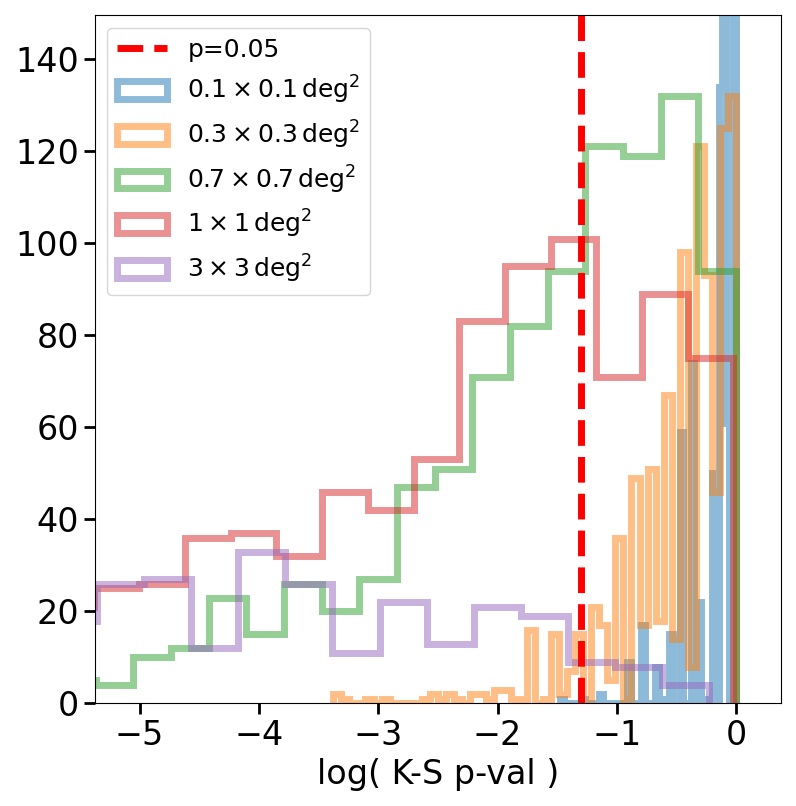}
    \caption{Goodness of the approximation with a lognormal distribution in terms of Kolmogorov-Smirnov p-value distributions ($\log_{10}$ scale). The vertical red line shows $p_{\rm KS}=0.05$. The Kolmogorov-Smirnov test is run over column density sets extracted from circular region of different areas in the $N_H$ all sky maps of \cite{2016A&A...594A.116H}.}
    \label{fig:KS_HI4PI}
\end{figure}
We draw real distributions from the all sky $N_H$ maps of \cite{2016A&A...594A.116H}. We consider this map representative of the physical column density at high Galactic latitudes $b>10 \deg$. We build column density sets extracted over circular regions of different dimension to test our assumption over different angular scales. For each extraction radius we draw 1000 sets centered at random positions in the sky at $b>10\deg$. From every set we obtain a probability of the set to be described by a lognormal distribution. We reject the hypothesis for p-values smaller than 0.05. The distributions of p-values obtained over sets at different angular scales are shown in Fig.~\ref{fig:KS_HI4PI}, for all p-values larger than $10^{-5}$.
The original $N_H$ map has a resolution of $\simeq 3.4'$. Larger angular scales then encompass a larger number of set elements. For instance, the $\sim 0.1 \,(1.0)\,\pi \deg^2$ extraction region encompass 30 (300) pixels. As shown in Fig.~\ref{fig:KS_HI4PI}, the column density distributions are generally consistent with a lognormal one up to $\sim 1\deg$ radius, with the corresponding distribution of p-values peaking at $p>0.05$. For larger regions, the real distribution starts to significantly deviate from lognormal. From a deeper inspection, this is due in most cases to a $N_H$ distribution resembling a sum of a few lognormal ones corresponding to portions at small scales encompassed within a single region.
The non-negligible width of such distributions makes a solution for the absorption computed from a single lognormal distribution still preferable to a correction by a single (average) value of column density.
Similarly, the same test performed over the Galactic center column density map shows consistency of the lognormal assumption up to a $6'$ radius angular scale, with larger regions deviating from a single lognormal.
The scale that limits the validity of the lognormal approximation has to depend in general on both the angular power spectrum of the column density fluctuations (and thus on the medium properties) as well as on the highest angular resolution used to sample the column density across the sky.

\subsection{Deviations from single column density absorption models}

So far we have demonstrated that: i) there are several cases of significantly broad column density distributions, produced by averaging over a large enough sky area; ii) in many of such cases, the distribution of column densities can be approximated with a lognormal one; iii) such an approximation leads to a more precise characterization of the absorption spectrum in terms of statistical power and accuracy of the best-fit values. To further demonstrate the importance of adopting a more complex model than only a single exponential function of the column density and also to assess when the latter is still a good enough approximation to the spectrum, we computed the discrepancy between the two methods, in terms of the quantities $A_\Theta$ and $\tilde{A}_\Theta$ already introduced in Sec.~\ref{sec:model}. Provided that the limit for small std$_\Theta$ of the \disnht\, model gives the \tbabs\, solution, the fractional (percent) deviation of $\tilde{A}_\Theta$ with respect to $A_\Theta$ is
\begin{equation}
    \left( \frac{A_\Theta}{\tilde{A}_\Theta}(E) -1 \right) \times 100
\end{equation}
and provides a mean to distinguish when the two models actually describe a significantly different absorption spectrum and to what extent.
\begin{figure}
    \centering
    \includegraphics[width=\linewidth]{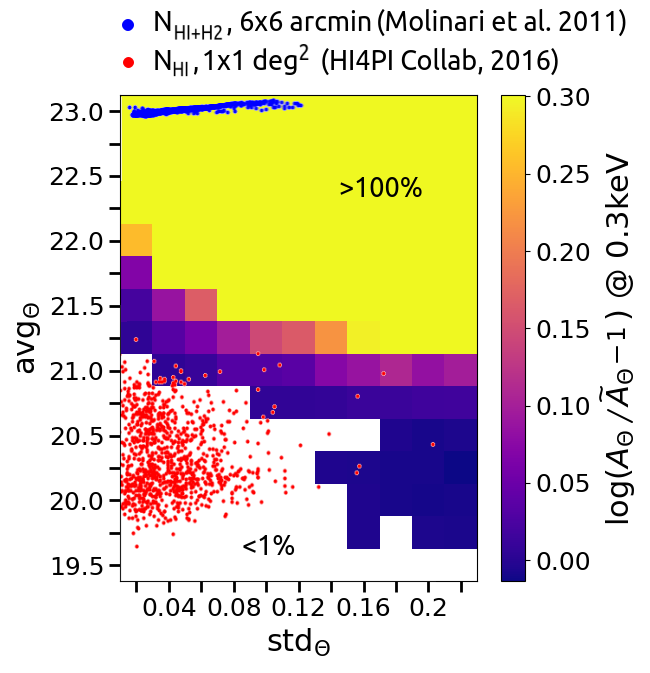}
    \caption{Percent deviation $(A_\Theta/ \tilde{A}_\Theta -1)\times 100$ (log scale) between the absorption at 0.3 keV produced by a lognormal distribution of column densities and a single-parameter absorption model (i.e. \tbabs\, $(N_H =\rm avg_\Theta)$), as a function of the free parameters avg$_\Theta$ and std$_\Theta$. Blue and red points show the scatter plot for distributions extracted respectively from $6' \times 6'$ random regions in the Galactic center column density map \citep{2011ApJ...735L..33M} and $1 \deg \times 1 \deg$ random region at Galactic latitudes $b>10 \deg$ from the HI column densities of \cite{2016A&A...594A.116H}.}
    \label{fig:correctionGrid_300eV_wp}
\end{figure}
We quantify the fractional deviation at energy 0.3 keV. At lower (higher) energies the models will deviate more (less). We study the deviation for a grid of parameters in the ranges avg$_\Theta \in [19.5; 23.0]$ using a 0.25 dex step and std$_\Theta \in [0.02; 0.22]$ using a 0.02 dex step. The results are shown in Fig~\ref{fig:correctionGrid_300eV_wp}.
Because of the exponential form, the difference between a negligible correction and an important correction is very large, even in log scale. We thus compress the color scale to highlight corrections between 1 and 30\% ($+100\%$ corresponding to a factor 2). A $\sim 1\%$ correction may in fact be statistically significant in a high signal-to-noise spectrum. White color regions of the parameter space indicate a negligible difference between the \disnht\, and \tbabs\, models. For such parameters the latter model is statistically favourable in the absence on priors on the distribution, since it requires only one (less) free parameter. 
We then plot in Fig.~\ref{fig:correctionGrid_300eV_wp} the points describing the column density distributions in the random regions extracted from the maps already considered in this work, for the upper bound angular scales where the lognormal approximation is verified, namely $1\deg \times 1 \deg$ for the \cite{2016A&A...594A.116H} map (red dots) and $6' \times 6'$ for the \cite{2011ApJ...735L..33M} Galactic center map (blue dots).
As we can see from Fig~\ref{fig:correctionGrid_300eV_wp}, for the majority of the high Galactic latitude regions the \tbabs\, model provides a good solution to the absorption. This is due to the column densities of absorbing material being on average lower than for lines of sight passing through the disk. Furthermore, the smoother absorbing medium at high Galactic latitudes distributes close around the mean value. In fact, we can see how the importance of the correction provided by the \disnht\, model increases toward the upper right corner of the plot (i.e. for increasing average column density, but also for increasing scatter around the mean). 
However, even at high latitude, there are regions that require a non-negligible correction provided by the \disnht\, model. The situation is drastically different for Galactic center observations and in general throughout the Galactic disk. In these regions both the average column density and the scatter around the mean increase of orders of magnitude. In these context, the large average value makes the absorption different than the \tbabs\, model even for narrower distributions and the correction factor is very large (i.e. $A_\Theta / \tilde{A}_\Theta >2$) already at 0.3 keV.

\section{Conclusions} \label{sec:conclusion}

We have presented a model for the absorption of soft X-ray spectra in the case that photons are collected from different lines of sight holding a broad distribution of column density values. After showing the efficacy of the model in retrieving the absorption properties of the spectrum, we proved that several cases exist for which it is relevant to consider it. In such cases the assumption of an absorption spectrum described by a single column density value breaks. We have proven that in such cases the lognormal distribution is a fair approximation to the physical one. 
The absorption model proposed is computed by assuming a constant layer of emission to be absorbed by a distribution of column densities. This is a rather simple assumption compared to a general case in which emission from a plasma is expected to deviate from homogeneity regardless of absorption. However, our goal is to focus on the modeling of the absorption of the spectrum, rather than on its emission. 
In general, whenever the source emission can be considered smooth with respect the angular scale of the absorption medium fluctuations, the \disnht\, model is valid and provides a better description of the absorbing medium with respect to a single-valued column density model. However, in case of a non smooth underlying emission, a correction for the column density distribution is still to be preferred to a simple average-value analysis of the absorption, in terms of result accuracy, for large enough column density distributions.
We expect the proposed model to be relevant for present and future generation of X-ray imaging telescopes furnished with survey capabilities and large fields of view, such as \erosita\, \citep{2012arXiv1209.3114M, 2021A&A...647A...1P} and {\it Athena} \citep{2013arXiv1306.2307N}.

The model is made available either through a {\tt python} code to compute the absorption spectrum given an arbitrary column density distribution, or precomputed under the assumption of a lognormal distribution both in {\tt ascii} and .fits tables ({\tt etable}, \xspec\, readable) formats\footnote{\url{http://www.brera.inaf.it/hotmilk/models.html}}. 

\section*{Acknowledgements}\label{acknowledgments}
We thank our anonymous reviewer for the helpful comments on the manuscript.
NL and GP acknowledge financial support from the European Research Council (ERC) under the European Union’s Horizon 2020 research and innovation program HotMilk (grant agreement No. [865637]). SB acknowledges financial support from the Italian Space Agency (grant 2017-12-H.0).

\vspace{-0.2cm}




\bibliographystyle{aa}
\bibliography{aa} 





\label{lastpage}
\end{document}